\documentclass[journal]{vgtc}                     

\onlineid{1641}

\vgtccategory{Research}

\vgtcpapertype{algorithm/technique}

\title{GhostUMAP2: Measuring and Analyzing \rd-Stability of UMAP}

\author{%
  Myeongwon Jung, 
  Takanori Fujiwara,
  and Jaemin Jo
}

\authorfooter{
  \item
  	Myeongwon Jung and Jaemin Jo are with Sungkyunkwan University.
  	E-mail: \{mw.jung@skku.edu, jmjo@skku.edu\}. Jaemin Jo is the corresponding author.
  \item
  	Takanori Fujiwara was with Linköping University. He is now with the University of Arizona.
  	E-mail: tfujiwara@ucdavis.edu
}

\abstract{%
Despite the widespread use of Uniform Manifold Approximation and Projection (UMAP), the impact of its stochastic optimization process on the results remains underexplored.
We observed that it often produces unstable results where the projections of data points are determined mostly by chance rather than reflecting neighboring structures.
To address this limitation, we introduce \rd-stability to UMAP: a framework that analyzes the stochastic positioning of data points in the projection space.
To assess how stochastic elements—specifically, initial projection positions and negative sampling—impact UMAP results, we introduce ``\textit{ghosts}'', or duplicates of data points representing potential positional variations due to stochasticity.
We define a data point's projection as \rd-stable if its ghosts perturbed within a circle of radius $r$ in the initial projection remain confined within a circle of radius $d$ for their final positions.
To efficiently compute the ghost projections, we develop an adaptive dropping scheme that reduces a runtime up to 60\% compared to an unoptimized baseline while maintaining approximately 90\% of unstable points.
We also present a visualization tool that supports the interactive exploration of the \rd-stability of data points.
Finally, we demonstrate the effectiveness of our framework by examining the stability of projections of real-world datasets and present usage guidelines for the effective use of our framework.}

\keywords{Dimensionality reduction, manifold learning, stochastic optimization, reliability, visualization, WebGPU}

\graphicspath{{figs/}{figures/}{pictures/}{images/}{./}} 

\usepackage{tabu}                      
\usepackage{booktabs}                  

\usepackage{xspace}
\usepackage{algorithm}
\usepackage[noend]{algpseudocode}
\usepackage{algorithmicx}
\usepackage{amsmath}
\usepackage{amssymb}
\usepackage{multirow}
\usepackage{booktabs}

\usepackage{listings} 
\usepackage[normalem]{ulem}

\newcommand{\rd}{$(r, d)$\xspace}

\newcommand{\gu}{Ghost\discretionary{-}{}{}UMAP\xspace}
\newcommand{\gutwo}{Ghost\discretionary{-}{}{}UMAP2\xspace}
\newcommand{\knn}{$k$NN\xspace}
\newcommand{\lazygen}{$lazy\_gen$\xspace}
\newcommand{\dropstart}{$drop\_start$\xspace}
\newcommand{\fone}{F1 score\xspace}
\newcommand{\sensitivity}{$sensitivity$\xspace}
\newcommand{\celegan}{\textit{C. elegans}\xspace}
\newcommand{\MNIST}{MNIST\xspace}
\newcommand{\agnews}{AG News\xspace}
\newcommand{\nepochs}{$n\_epochs$\xspace}

\newcommand{\ghostexplorer}{GhostExplorer\xspace}
\newcommand{\graph}{$\mathcal{W}$\xspace}
\algnewcommand{\IIf}[1]{\State\algorithmicif\ #1\ \algorithmicthen}
\algnewcommand{\EndIIf}{\unskip\ \algorithmicend\ \algorithmicif}

\newif\ifshowmwdelete
\showmwdeletefalse

\newif\ifshowdel
\showdelfalse

\newcommand{\mw}[1]{\textcolor{black}{#1}}

\newcommand{\edit}[1]{\textcolor{black}{#1}}

\newcommand{\del}[1]{\ifshowdel \textcolor{red}{\sout{#1}} \else \ignorespaces \fi}

\usepackage{tabu}
\usepackage{booktabs} 
\usepackage{enumitem}
\usepackage{kotex}

\usepackage{multicol}
\usepackage{mathptmx}                  

\usepackage{url}
\begin{document}

\crefname{algorithm}{Alg.}{Algs.}
\Crefname{algorithm}{Algorithm}{Algorithms}

\crefname{listing}{Code}{Codes}
\Crefname{listing}{Code}{Codes}




\firstsection{Introduction}

\maketitle

We introduce \rd-stability to Uniform Manifold Approximation and Projection (UMAP), a framework designed to evaluate the stability of UMAP projections.
UMAP is among the most popular dimensionality reduction (DR) techniques, distinguished by its computational efficiency.
To achieve efficiency, the official \del{UMAP} implementation~\cite{mcinnes2018umap} incorporates various acceleration techniques, such as \del{stochastic gradient descent and} negative sampling~\cite{mikolov2013distributed, tang2015line, tang2016visualizing}. However, these techniques introduce stochasticity, which adds variability to the results. 
This variability can obscure the distinction between a data point's projection influenced by stochasticity and one that is stable, thereby complicating robust interpretation.

We identified three sources of stochasticity in the UMAP implementation.
First, UMAP generates an initial embedding using a stochastic process, such as Spectral Embedding~\cite{belkin2001laplacian}, and adds small random noise to the embedding to avoid local minima for optimization. 
Second, UMAP uses negative sampling where small random samples are drawn at each iteration to simulate repulsive forces among points, instead of using all pairs of points in the dataset.
Third, UMAP exploits parallelization \mw{without synchronization}, facilitated by the Python Numba library~\cite{lam2015numba}, \del{without synchronization} which can result in race conditions.

\del{
Our \rd-stability is designed to gauge the variability resulting from the first two sources by computing the potential positional variations of each point.
Specifically, we consider a circle of radius $r$ around a target point in the initial embedding and analyze how its shape transforms during UMAP's non-linear optimization process.
If the circle remains approximately circular with a reduced radius, it suggests that the target point resided in a relatively stable region in the initial projection, making its position in the final projection more reliable in terms of robustness to the stochasticity.
Conversely, if the circle undergoes significant fragmentation, it may indicate that the target point and its initial projection are more vulnerable to variability, signaling instability in the final projection.
}
\mw{In this work, we define \rd-stability as the stability of a point's projection with respect to the variability arising from the first two sources.
To assess this, we place hypothetical points, referred to as \textit{ghosts}, around a target point in the initial embedding and observe how their projections evolve through UMAP’s optimization process.
Due to slight differences in their initial positions and the stochasticity of negative sampling, these projections may diverge from one another.
If the ghost projections converge near the target point’s projection, we consider the projection to be stable.
Otherwise, if they fail to converge, it may indicate that the target point is vulnerable to variability, signaling instability in the final projection.} 
It is worth noting that the instability we discuss here differs from distortion, which inherently arises when reducing high-dimensional data to lower-dimensional embeddings.
Rather, our focus is specifically on instability driven by the stochastic algorithms, aiming to identify points sensitive to this randomness.


We present \gutwo, an efficient DR algorithm to evaluate the \rd-stability of points projected by UMAP.
\mw{To define ghosts and compute their projections efficiently,} we sample points \del{use discrete samples} from \mw{a circle centered at the target point. These ghosts share} the identical high-dimensional representation with the target but have perturbed initial positions. 
We asymmetrically simulate the forces between the original points and ghosts; the ghosts are invisible to the original points. Therefore, the projection of the original points is not affected by our technique.
In contrast, we optimize the ghosts with different negative samples, allowing them to diverge from the original points.
Finally, the \mw{distance to the farthest ghost projection} after optimization is used as a measure of the target point’s stability.

This work extends a short paper~\cite{jung2024ghostumap} presented at IEEE VIS 2024 with three significant advancements over the previous contribution.
First, we provide a formal definition of \rd-stability and apply it to \del{a widely used DR technique, } UMAP.
While doing so, we expand our analysis by accounting for the variability introduced from the random initial embedding in addition to that from negative sampling; indeed, our previous work can be seen as a special case of \gutwo with $r=0$.
Second, we introduce an adaptive dropping scheme to accelerate the computation of \rd-stability, and achieve approximately a 30\% speedup over our previous successive halving scheme and up to 60\% speedup compared to an unoptimized baseline.
Finally, we present an interactive visualization tool, \ghostexplorer, for analyzing \rd-stability of a projection \del{, which is accelerated by the WebGPU technology.
We also provide} \mw{and guidelines on hyperparameter settings and stability interpretation.}
\section{Related Work}

\subsection{Uniform\,Manifold\,Approximation\,and\,Projection\,(UMAP)}
\label{sect:umap}

To explain the details of \gutwo, we provide a concise introduction to the core components of UMAP from a computational perspective.
The UMAP algorithm consists of two main phases: graph construction and layout optimization.

\textbf{Graph Construction.}
The graph construction phase aims to construct a topological representation of high-dimensional data by building a weighted $k$-Nearest Neighbor ($k$NN) graph which combines local fuzzy simplicial sets.
Formally, let $X = \{x_1, \ldots, x_N\}$ be the given high-dimensional dataset, with a distance metric $d: X \times X \rightarrow \mathbb{R}_{\geq 0}$.
For each data point $x_i$, the set of $k$-nearest neighbors, $\mathcal{N}_i$, is determined regarding $d$.
Then the weight of a directed edge from $x_i$ to $x_j$, representing the probability of the edge's existence, is computed as:
\begin{equation} \label{eq:v}
    v_{j|i} = \exp({-\max(0, d(x_{i},x_{j})-\rho_{i})} / \sigma_{i}).
\end{equation}
Here $\rho_i$ and $\sigma_i$ are local connectivity parameters, ensuring the assumption that data points are uniformly distributed over a manifold in high-dimensional space.
The weight of an undirected edge between $x_i$ and $x_j$, denoted $v_{ij}$, is then computed as the combined probability of at least one directed edge existing: $v_{ij} = (v_{j|i} + v_{i|j}) - v_{j|i} \cdot v_{i|j}$.

\textbf{Layout Optimization.}
The layout optimization phase maps the topological graph representation onto a low-dimensional space while preserving its structural properties.
This is achieved using a force-directed graph layout algorithm that applies attractive and repulsive forces between points in the low-dimensional space.
These forces are derived from gradients of the edge-wise cross-entropy loss, which measures the difference between the high-dimensional and low-dimensional weighted graphs.

The low-dimensional projection $Y = \{y_1, \cdots, y_N \}$ is initialized using spectral embedding.
The weight of an edge between points $y_i$ and $y_j$ is defined as:
\begin{equation} \label{eq:w}
w_{ij} = (1 + a|| y_{i} - y_{j} ||_{2}^{2b})^{-1}, 
\end{equation}
where $a$ and $b$ are user-specified positive parameters.

The cross-entropy loss, minimized during optimization, is given by:
\begin{equation} \label{eq:ce}
    CE = \sum_{i \neq j} \left[v_{ij} \cdot \log\left(\frac{v_{ij}}{w_{ij}}\right) - (1-v_{ij}) \cdot \log\left(\frac{(1-v_{ij})}{(1-w_{ij})} \right)\right].
\end{equation}
From this loss function, the attractive and repulsive forces acting on $y_i$ due to $y_j$ are computed as:
\begin{equation}\label{eq:att}
    F_{att}(y_i,y_j,v_{ij}) = \frac{-2abd_{ij}^{2(b-1)}}{1 + d_{ij}^2} v_{ij} (y_i - y_j)
\end{equation}
\begin{equation}\label{eq:rep}
    F_{rep}(y_i,y_j,v_{ij}) = \frac{2b}{(\epsilon + d_{ij}^2)(1 + ad_{ij}^{2b})}(1 - v_{ij})(y_i - y_j)
\end{equation}   
where $d_{ij}=\|y_i - y_j\|_2$ and $\epsilon$ is a small constant to prevent division by zero.
The optimization process iteratively applies these forces to refine the projection.
The attractive force is applied between pairs of points connected by an edge in the topological graph, pulling them closer together.
In contrast, the repulsive force theoretically needs to be computed between each point and all other points, which is computationally expensive.

To reduce this complexity, UMAP employs negative sampling, where for each point $i$, a subset of \textit{n\_negative\_samples} points ($n\_negative\_samples = 5$ by default) is randomly sampled.
The repulsive force is applied only between $i$ and the sampled points, substantially improving computational efficiency.
However, this stochastic approach introduces randomness to the optimization process, contributing to the inherent variability in UMAP's results.

\subsection{Distortion and Instability of DR}

To the best of our knowledge, this work is the first to measure and analyze the instability caused by the stochastic nature of DR techniques. In contrast, prior research has primarily focused on the distortions introduced by DR techniques when mapping high-dimensional data to low-dimensional spaces. These distortions have been studied extensively at various levels, including local~\cite{lee2009quality, venna2006local, fujiwara2023feature}, cluster~\cite{jeon2021measuring, paulovich2008least, sips2009selecting}, and global levels~\cite{hinton2002stochastic, kruskal1964multidimensional}, resulting in measures that quantify the amount of the distortion. Such measures have 
been used to assess the quality of DR results and to compare different techniques~\cite{espadoto2019toward, jeon2023zadu, nonato2018multidimensional}.
Visual interfaces have also been developed to facilitate the analysis and interpretation of the distortions~\cite{lespinats2011checkviz, stahnke2015probing}.

However, it is important to note that these studies on distortion focus on the relationship between a DR result and the original high-dimensional space, without accounting for the variability introduced during the generation of the result by the DR technique itself.
The most closely related work to ours is by Jung et al.~\cite{jung2023projection}, where the authors compared multiple $t$-SNE projections to identify stable clusters across multiple runs.
\del{In contrast, our \gutwo approach measures the instability of projections during UMAP, providing a framework for analyzing the stochasticity of UMAP more efficiently.}
\mw{In contrast, our approach measures the instability of a projection in a single run, which is computationally effective.}

\begin{figure}[t]
    \centering
    \includegraphics[width=0.7\linewidth]{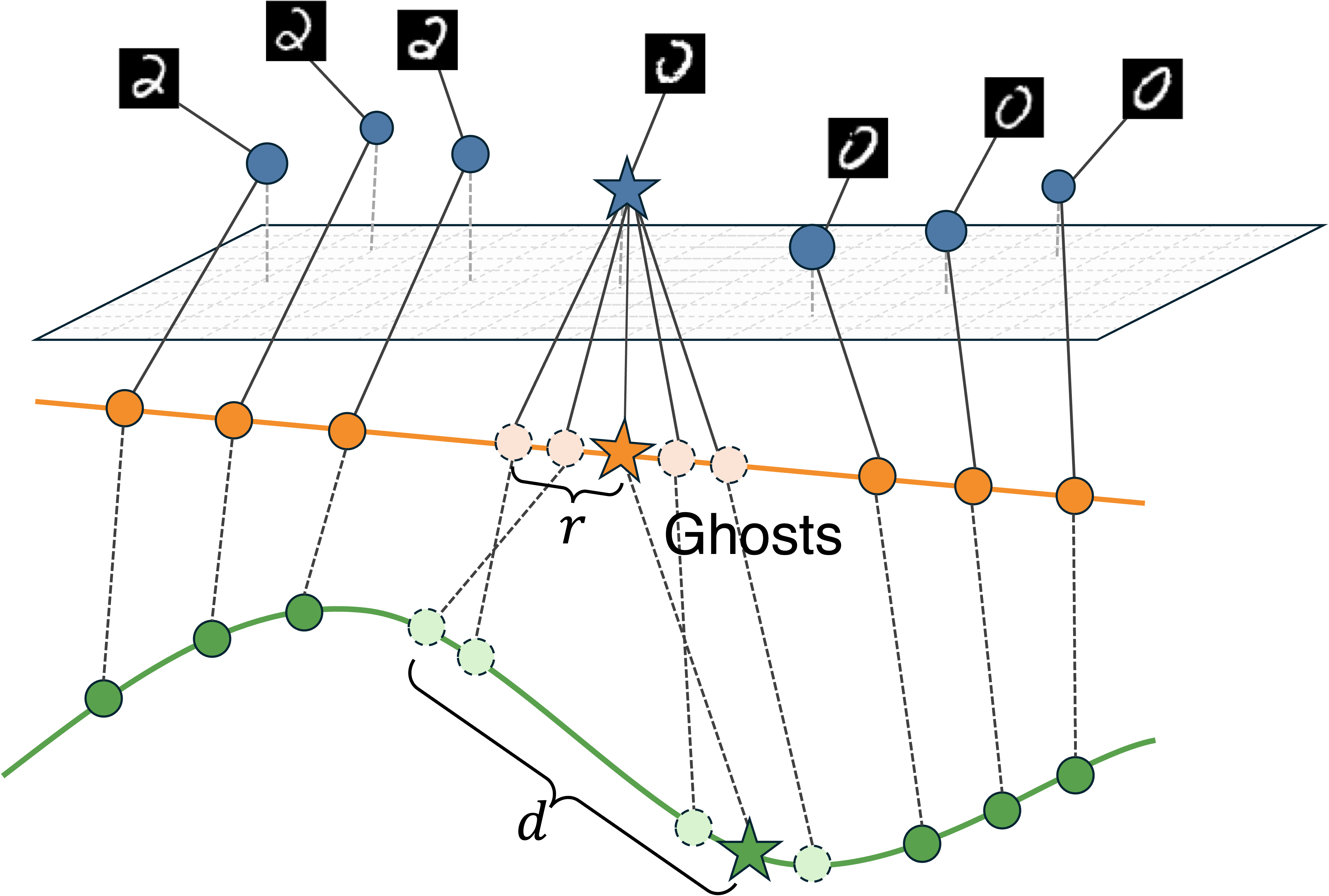}
    \caption{
    A motivating example illustrating the \rd-stability framework.
    The figure shows the original high-dimensional data points (colored blue), the initial 1D projection generated by PCA (orange), and the final 1D projection (green).
    The starred point is vulnerable to stochasticity as its image is ambiguous between two classes (digits zero and two).
    We attach ghosts to its initial projection (orange dashed line) and jointly optimize the original and ghost projections.
    As a result, the distribution of ghost projections can highlight the stochastic instability of the point.
    }
    \label{fig:manifold}
    \vspace{-5mm}
\end{figure} 

\section{\gutwo}

We introduce \gutwo, incorporating the concept of \rd-stability into UMAP.
With a motivating example, we define ghost projections to measure the \rd-stability and elaborate on the layout optimization process for ghost projections.
We further present an adaptive dropping scheme to reduce the computational overhead introduced by the ghost projections.
We then present a visualization tool, \ghostexplorer, to analyze the results of \gutwo.

\subsection{A Motivating Example}
\label{sect:motivation}

The layout optimization process of UMAP starts with an initial projection created by simpler DR methods, such as Principal Component Analysis (PCA)~\cite{hotelling1933analysis} or Spectral Embedding~\cite{lam2015numba}. 
\cref{fig:manifold} illustrates the relationships between the original data (colored blue), initial projection (orange), and final projection (green). 
For illustration, consider a scenario where a 1D manifold in 3D space, conceptualized as a \mw{hypothetical} curved line \del{connecting blue points} \mw{formed by the blue points} in \cref{fig:manifold}, is projected onto a 1D space.
In this example, each original high-dimensional data point corresponds to a hand-written digit 0 or 2.

During the optimization process, the initial projection (orange line in \cref{fig:manifold}) is iteratively annealed to the final projection (the green line) to best describe the manifold in the high-dimensional space where the goodness of fit is computed as UMAP's objective function (\cref{eq:ce}).
However, since the optimization process is stochastic, the final projection is subject to variability.
For example, the starred point in \cref{fig:manifold} is pushed more toward the cluster of digit 0 by chance, misleading the user into concluding the starred point is clearly different from digit 2.
Our framework would identify that this starred point is not \rd-stable.

To specify data points prone to variability, we introduce clones of a target point, called ghosts, which are shown as dotted orange circles in \cref{fig:manifold}.
In the initial projection, we randomly place ghosts around the target point within a distance $r$ where $r$ is a user-specified hyperparameter.
In practice, UMAP is often used to generate a 2D projection and we place ghosts within a $r$-radius circle centered at the target point.
While these ghosts have different projection positions, they still have the same high-dimensional vector as the target point.
The ghosts can be viewed as alternative projections of the target point with different starting positions created by adding perturbations.
We then jointly optimize the projection positions of the ghosts as well as the target point (detailed in \cref{sect:optimization}).
The result is shown as dotted green circles in \cref{fig:manifold}.
We can see that two ghosts are projected closer to the cluster of digit 2 and the other two closer to digit 0, implying the original projection (the green star) is highly subject to variability, and thus, its position should be interpreted with care.

\subsection{\rd-Stability of a Point in UMAP}

We formulate the \rd-stability of a point in a UMAP projection.
Let $X = \{x_i\ |\ i\in[1, ..., N] \}$) be a set of high-dimensional input vectors where $x_i \in \mathbb{R}^D$ ($D$: the \del{nubmer}\mw{number} of dimensions) and $Y = \{y_i\ |\ i\in[1, ..., N]\}$ be a set of initial projections of $X$ where $y_i \in \mathbb{R}^2$.
The optimization process of UMAP can be seen as a function that maps the initial projection $Y$ to the final projection $Y^\prime$, i.e., $Opt: (Y, \mathcal{W}, \mathcal{H}) \mapsto Y^\prime$ where \graph is the weighted \knn graph of $X$ (refer to \cref{sect:umap}) and $\mathcal{H}$ is a set of UMAP's hyperparameters.
We assume both $Y$ and $Y'$ are normalized into a range $[[0, 1], [0, 1]]$

In \gutwo, each point $x_i$ has $M$ ghost projections, $g_{i_k} \in \mathbb{R}^2$ ($k\in[1, ..., M]$).
Each ghost's projection is initialized to a random position in a circle of radius $r$ around the initial projection of the target $y_i$:
\begin{equation}
\label{eq:sampling}
    g_{i_k} \gets SampleCircle(y_i, r)
\end{equation}
\mw{We chose a circle over other shapes for simplicity. While one could sample ghost projections exactly at a distance $r$ from the target, i.e., from the circumference of a circle, we opted to sample from within the circle’s interior for two reasons. First, our intention was to approximate a ball rather than a shell. Second, this choice aligns with UMAP’s initialization, where each point's projection is given a small random perturbation sampled from within a bounded range.}

Let $G = \{g_{i_k}|\ i\in[1, ..., N],\ k\in[1,\dots,M]\}$ be the set of the ghost projections of all data points in $X$. The optimization of \gutwo is a function that jointly optimizes the initial projections of original points ($y_i \in Y$) and ghosts ($g_{i_k} \in G$):
\begin{equation}
    LayoutOptimization: (Y, G, \mathcal{W},\mathcal{H}) \mapsto (Y^\prime, G^\prime)
\end{equation}
where $G'$ is the final projection of the ghosts.
We let $d_i$ denote the maximum Euclidean distance between the final projection positions of the target point, $y^\prime_i$, and its ghost, $g^\prime_{i_k}$:
\begin{equation}
\label{eq:di}
   d_i := \max_{k} \|y'_i - g'_{i_k}\|_2 
\end{equation}
We define $x_i$ \rd-stable if $d_i \leq d$ with a user-defined threshold $d$. 


The hyperparameter $r$ can be seen as ``stability loads'' given to each point.
Increasing $r$ will make fewer points be \rd-stable for the same $d$.
In contrast, $d$ can be adjusted after the optimization without recomputation and determines how generous we are about the instability.
Decreasing $d$ will make fewer points be \rd-stable.
In practice, the user may want to decrease $d$ as much as possible while keeping a sufficient number of points remaining \rd-stable.
Then, the user can refer to these stable points to interpret the projection results or the unstable points to investigate the cause of instability (refer to our use cases in \cref{sect:use_cases}).
Both $r$ and $d$ are in the range $[0, 1]$ as they are defined in normalized coordinates.

Suppose a perfectly deterministic DR algorithm that projects $x_i$ to a fixed point $y^\prime_i$ regardless of its initial projection $y_i$ without any negative sampling.
In this ideal case, $x_i$ is $(1, 0)$-stable as there is no variability in $y^\prime_i$ even though we impose an infinite amount of perturbation on $y_i$.
On the other extreme, suppose a fully random algorithm that maps $x_i$ to a random point.
In this worst possible case, $x_i$ is $(0, 1)$-stable as there is no bound in the projection  $y^\prime_i$ even though we impose no perturbation.

\subsection{Ghost Generation and Joint Optimization}\label{sect:optimization}

The optimization process in \gutwo consists of three phases: ghost generation, layout optimization, and adaptive ghost dropping (\cref{alg:main}).
This subsection focuses on the first two phases, ghost generation and layout optimization.
In the ghost generation phase, we randomly generate $M$ ghosts by following \cref{eq:sampling}.

We can attach the ghosts not to the very initial projection but after running a few layout optimization iterations of UMAP, which we call lazy generation.
Such deferred generation can have three benefits:
First, during the early epochs of layout optimization, the UMAP's learning rate $\alpha$ is at its peak, resulting in rapid changes to the projection.
These early decisions often dominate the process, making it challenging to evaluate the influence of randomness in later stages.
Second, the initial projection is often much different from the final projection, and thus, measuring the stability using the initial projection may be less meaningful. After a few optimization iterations, the projection better fits the manifold we want to approximate, which can be a more meaningful starting point.
Third, delaying ghost generation reduces computational overhead as we can start optimization in a later epoch, thereby reducing the runtime.
In our implementation, this is controlled by a hyperparameter, $lazy\_gen\ \in\ [0, 1)$, which is multiplied to \nepochs to determine at which epoch we generate ghosts.

\begin{algorithm}[t]
\caption{GhostUMAP Layout Optimization}\label{alg:main}
\begin{algorithmic}[1]
\scriptsize
\Function{LayoutOptimization}{$Y, G, \mathcal{W}, \mathcal{H}$}
    \State $\alpha \gets 1.0$  ~~~~~~(note: $\alpha$ is a learning rate)
    \State {$e_{gen} \gets \lceil n\_epochs \times lazy\_generation 
    \rceil$}
    \State {$e_{drop} \gets \lceil n\_epochs \times drop\_start 
    \rceil$}
    \State $G \gets \textbf{None}$
    \State $D_i \gets 0$ for all $i\in[1,\dots,N]$
    \For{$e \gets 1$ to $n\_epochs$}
        \If{$e \geq e_{gen}$ and $G$ is None}
            \State $G \gets \textsc{SampleCircle}(Y, r)$
        \EndIf
        \State \Call{OptimizeOriginalLayout}{$Y, \mathcal{W}$}
        \State \Call{OptimizeGhostLayout}{$Y, G, \mathcal{W}$} 
        \State $D \gets$ \Call{UpdateDistances}{$Y, G, D$}
        \If{$e \geq e_{drop}$}
            \State \Call{DropGhosts}{$G, D$} 
        \EndIf
        
        \State $\alpha \gets 1.0 - e /n\_epochs$
    \EndFor
    \State \Return {$Y, G$}
\EndFunction
\end{algorithmic}
\end{algorithm}



\begin{algorithm}[t]
\caption{Layout Optimization for Original Projections}\label{alg:olo}
\begin{algorithmic}[1]
\scriptsize
\Function{OptimizeOriginalLayout}{$Y,\ \mathcal{W}$}
\ForAll{$ ([i,\ j], v_{ij}) \in \mathcal{W}$}
    \State $f_{att} \gets F_{att}(y_i, y_j, v_{ij})$
    \State ($y_i, y_j) \gets (y_i + \alpha \cdot f_{att}, y_j - \alpha \cdot f_{att})$

    \For{$n \gets 1,\dots,n\_negative\_samples$}
        \State $l \gets \text{random sample from } \{1, \dots, N\}\backslash \{i\} $
        \State $y_i \gets y_i + \alpha \cdot F_{rep}(y_i, y_l, v_{il})$
    \EndFor
\EndFor
\EndFunction
\end{algorithmic}
\end{algorithm}

\begin{algorithm}[t]
\caption{Layout Optimization for Ghost Projections}\label{alg:glo}
\begin{algorithmic}[1]
\scriptsize
\Function{OptimizeGhostLayout}{$Y, G, \mathcal{W}$}
    \State \textbf{if} $G$ is None \textbf{then} \textbf{return}

    \ForAll{$([i, j], v_{ij}) \in \mathcal{W}$}
        \ForAll{$k \gets 1, \dots, n\_ghosts$}
            \State $g_{i_k} \gets g_{i_k} + \alpha \cdot F_{att}(g_{i_k}, y_j, v_{ij})$
            \For{$n \gets 1, \dots, n\_negative\_samples$}
                \State $l \gets \text{random sample from } \{1, \dots, N\} \backslash \{i\}$
                \State $g_{i_k} \gets g_{i_k} + \alpha \cdot F_{rep}(g_{i_k}, y_l, v_{il})$
            \EndFor
        \EndFor
    \EndFor
\EndFunction
\end{algorithmic}
\end{algorithm}

\begin{figure*}[t!]
    \centering
    \includegraphics[width=\linewidth]{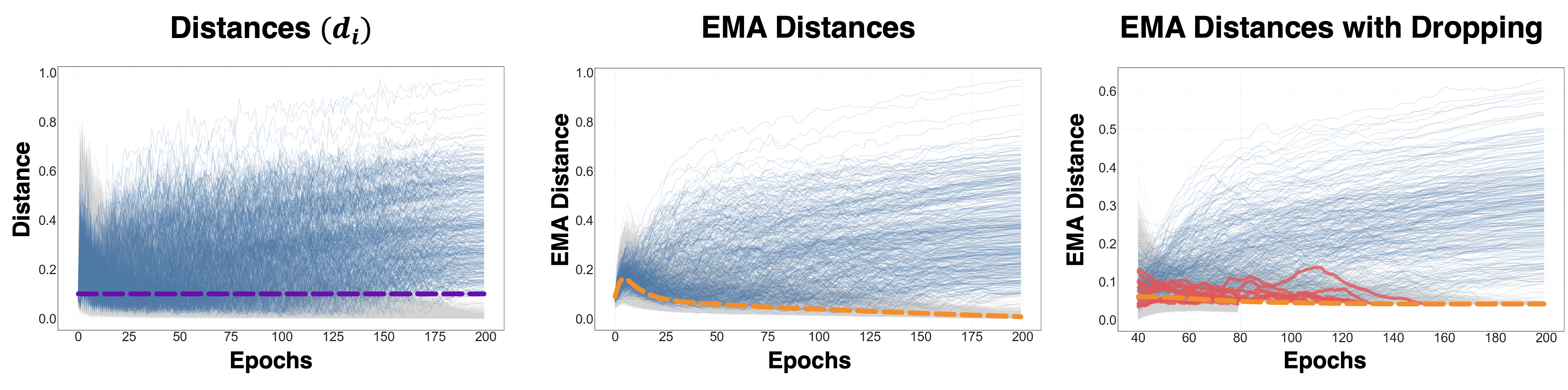}
    \caption{
    Distance changes of stable (gray) and unstable (blue) projections by optimization epochs.
    \textbf{Left}: We choose points with $d_i \geq 0.1$ in the final projection as unstable points, whose threshold ($d=0.1$) is shown as the purple dashed line. 
    \textbf{Middle}: We use Exponential Moving Average (EMA) to mitigate the fluctuation and set the dropping threshold as the mean of the averaged distances (orange dashed line).
    \textbf{Right}: In practice, we freeze the $D_i$ of dropped points (orange dashed line). The false positives, i.e., the points that are unstable but dropped mistakenly, are shown as red lines. The dropping operation started at epoch 80.
    }
    \label{fig:dropping}
    \vspace{-5mm}
\end{figure*}
 
We set three requirements for layout optimization.
First, the optimization of ghost projections must not interfere with the original projections.
This ensures that original points are projected as if the ghosts were absent, thereby producing projections identical to those of the original UMAP.
Second, ghost projections must accurately represent their corresponding projections of the original points. 
To achieve this, when generating ghosts, we add the perturbation only to the projection positions, and all ghosts still have the same high-dimensional vector, $x_i$, as their target point.
Third, separate negative samples should be used for the original and ghost projections to evaluate the impact of negative sampling on the optimization process.

The joint optimization of original and ghost projections \del{are}\mw{is} outlined in Lines \del{11 and 12}\mw{10 and 11} in \cref{alg:main}.
We first optimize the original projections $Y$ as with the standard UMAP (\cref{alg:olo}).
Note that, to ensure no influence from the ghost on the original projections, 1) the weighted \knn graph, $\mathcal{W}$, is generated without using ghosts and 2) ghosts are not chosen by negative sampling.

The optimization process for ghost projections is detailed in \cref{alg:glo}. We introduce two modifications when compared with \cref{alg:olo}. 
First, the attractive and repulsive forces are computed with and applied to a ghost, $g_{i_k}$, instead of $y_i$ (Lines 5 and 8 in \cref{alg:glo}).
This ensures that ghost projections serve as alternatives to the original projections. 
Second, we negative-sample only the original projections again (i.e., ghosts are not used as negative samples), which would be different from the negative samples used for the original point (Lines \del{6-9}\mw{6-8} in \cref{alg:glo}).
\del{This negative sampling is to account for the impact of varying negative samples.}
\mw{This allows each ghost projection to model the outcome that would result if different negative samples were assigned to its corresponding original projection.}

The joint optimization of the original and ghost projections can be summarized at a high level as follows: the original projections are optimized in a way that they are unaware of ghosts. Each ghost is also unaware of other ghosts but is optimized as if it were the original projection of a point.

\begin{algorithm}[tb] \caption{Adaptive Dropping Scheme} \label{alg:dropping}
\begin{algorithmic}[1]
\scriptsize
\Function{UpdateDistances}{$Y$, $G$, $D$}
    \IIf{$G$ is \textbf{None}} \Return \EndIIf
    \State {$j \gets \lceil M \times sensitivity \rceil$}
    \ForAll{$i\in \{\text{id}(g_i) |\ g_i \in G$\}}
    \State $d_i \gets$ distance from $y_i$ to the $j$-th farthest ghost $g_{i_k}$ 
    \State $D_i \gets \beta \cdot d_i + (1 - \beta) \cdot D_i$ \EndFor
\State \Return {$D$}
\EndFunction
\State
\Function{DropGhosts}{$G$, $D$}
    \State Compute threshold: $\tau \gets$ mean of $D$
    \State Keep unstable points in $G$: $G \gets \{ g_i \in G \mid D_i \geq \tau \}$
\EndFunction
\end{algorithmic}
\end{algorithm}

\subsection{Adaptive Dropping Scheme}

Jointly optimizing the ghost projections with the original projection requires extra time to compute.
Roughly speaking, having $M$ ghosts (by default, $M=16$) is equivalent to optimizing $M$ times more data points than the original ones, slowing down the process significantly.
To reduce such performance overhead, in the previous work~\cite{jung2024ghostumap}, we introduced a successive halving (hereafter, halving) scheme.
The halving scheme sorts data points by the positional variance in the original and ghost projections and performs a halving operation that discards the ghosts of half of the points with the lowest variance (i.e., stable points).

One significant constraint of the halving scheme is that the user should determine epochs at which they want to perform the halving operation, i.e., a halving schedule, before they see how the optimization proceeds.
This constraint leads to two drawbacks.
First, a fixed schedule may miss the opportunity to drop ghosts of the stable points as early as possible since the dropping needs to wait until the next halving epoch is reached.
Second, this scheme only allows for a fixed ratio of data points' ghosts to be dropped, merely guaranteeing a fixed amount of speedups. For example, if we schedule three halving operations, only 87.5\% points will be dropped eventually, even though more points can be found to be sufficiently stable.

To address the limitation of the halving scheme, we introduce an adaptive dropping scheme that identifies stable points as early as possible, as outlined in \cref{alg:dropping}.
Rather than dropping a fixed number of ghost projections at a certain epoch, our new scheme constantly checks if a point is stable enough by monitoring $d_i$---the maximum distance between its original projection and ghosts (\cref{eq:di}).
At the beginning, $d_i \leq r$ because all ghosts are in a $r$-radius circle centered at the target point.
However, as the optimization proceeds, $d_i$ can fluctuate.
We are interested in identifying points whose $d_i$ is maintained less than a certain threshold and dropping their ghosts.
In contrast, points with large $d_i$ should avoid dropping their ghosts as they are of interest.


To establish a dropping criterion, we examined how $d_i$ of a stable or unstable point changes during optimization, as shown in \cref{fig:dropping}.
In the left chart, the gray and blue lines depict $d_i$ of stable and unstable points, respectively.
For illustration, we chose points with $d_i \geq 0.1$ in the final projection as unstable points. We highlighted the threshold as a purple dashed line.
Note that although there is a fluctuation, $d_i$ of a stable point (gray) gradually converges, while $d_i$ of an unstable point does not.
We leverage this observation of the convergence to identify stable points.
Instead of directly referring to $d_i$, we compute the smoothed distance $D_i$ by applying exponential moving average (EMA)~\cite{ema} to $d_i$ with a smoothing factor $\beta$ (0.2 by default), as shown on Line 6 of \cref{alg:dropping}.
This smoothing reduces volatility, as shown in the middle plot of \cref{fig:dropping}, allowing a clearer separation of stable and unstable points.
For each iteration, we use the mean of $D_i$ as a dropping threshold, $\tau\ =\ mean(D_i)$, and drop the ghosts of the points whose smoothed distance $D_i$ is less than $\tau$ (Lines \del{14--15}\mw{10--11} in \cref{alg:dropping}). 
$\tau$ over iterations is shown as the orange dashed line in the middle chart of \cref{fig:dropping}.
\mw{
One can consider using medians instead of means. However, doing so is essentially similar to the halving scheme. Furthermore, as medians are more robust than means, we found them to be overly conservative, often retaining even very stable points.
}

One practical concern on $\tau$ is the inflation of the threshold as stable points are dropped.
For example, in practice, about half of the points are dropped in the first dropping operation as $\tau$ is set to $mean(D_i)$.
In the next iteration, if we compute $\tau$ again only with the remaining half of the points, i.e., points with higher $D_i$, $\tau$ will rapidly increase.
We observed that such an inflation of $\tau$ degrades the quality of dropping by mistakenly discarding unstable points (refer to the red lines in \cref{fig:dropping}-Right). 
To prevent this issue, instead of eliminating $D_i$ of the points with dropped ghosts, we include their $D_i$ when computing $\tau$ while keeping their $D_i$ as the same value even after the ghost dropping.

Another concern is that since $d_i$ is determined by the farthest ghost projection, it is sensitive to outliers.
Indeed, we observed that certain negative samples push the ghost projections far away from the original projection by chance.
To prevent such influence by outliers, we introduce a hyperparameter $sensitivity \in [0, 1]$ (0.9 by default) that determines the percentile of the ghost projection distances to consider.
If it is set to 1.0, we consider the distance to the farthest ghost projection.
When set to 0.9, we use the 90th percentile of the distances to ghosts.

The last question is about when to start the dropping operation. 
One can start dropping operations immediately after ghosts are initialized.
This may allow for reducing the runtime by dropping the ghosts as early as possible but at the risk of mistakenly dropping the ghosts of unstable points since the projection is premature. 
To address this, we start the dropping operation after a certain portion of epochs is done, specified by a hyperparameter, $drop\_start \in [0, 1]$ (0.4 by default).
We will investigate the effect of these hyperparameters in \cref{sect:eff_hparam}.

The \gutwo implementation is written in Python and distributed as a Python package (\texttt{ghostumap}), making it easily installable via the Python Package Index (\texttt{pip}) or Anaconda.
The package employs an API consistent with UMAP, such as \texttt{gmap.fit\_transform} in \cref{code:ghostumap_example}, with additional hyperparameters for our technique, such as the number of ghosts.
To enhance efficiency, we also utilized the Numba package \del{, which is similar to UMAP} \mw{in the same way as UMAP}.

\lstdefinestyle{pythonstyle}{
    language=Python,
    basicstyle=\ttfamily\small,
    keywordstyle=\color{blue}\bfseries,          
    stringstyle=\color{red},            
    commentstyle=\color{gray},          
    numbers=left,
    numberstyle=\tiny\color{gray},
    stepnumber=1,
    frame=single,
    breaklines=true,
    breakindent=1.26em,
    backgroundcolor=\color{white},
    moredelim=**[is][\color{teal}]{@}{@}, 
    moredelim=**[is][\color{orange}]{!}{!},
    moredelim=**[is][\color{YellowGreen}]{*}{*},
}

\renewcommand{\lstlistingname}{Code}

\begin{lstlisting}[style=pythonstyle, caption={The Python Interface of GhostUMAP2}, label={code:ghostumap_example}, float=tp, linewidth=\linewidth]
from @ghostumap@ import @GhostUMAP2@
from @data@ import !load_data!

X, y, legend = !load_data!()
@gmap@ = @GhostUMAP2@()
O, G, survived = @gmap@.!fit_transform!(X, n_ghosts=16, r=0.1)

# N: the number of data points, i.e., X.shape[0]
# O: original projections of shape (N, 2)
# G: ghost projections of shape (N, n_ghosts, 2)
# survived: a bitmask of length N indicating if point i has not been dropped.

unstable = @gmap@.!get_unstable!(d=0.1)
# unstable: a bitmask of length N indicating if point i is unstable with respect to d.

@gmap@.!visualize!(label=y, legend=legend)
\end{lstlisting}

\subsection{\ghostexplorer: A Visualization Tool for \gutwo}

\begin{figure}
    \centering
    \includegraphics[width=\linewidth]{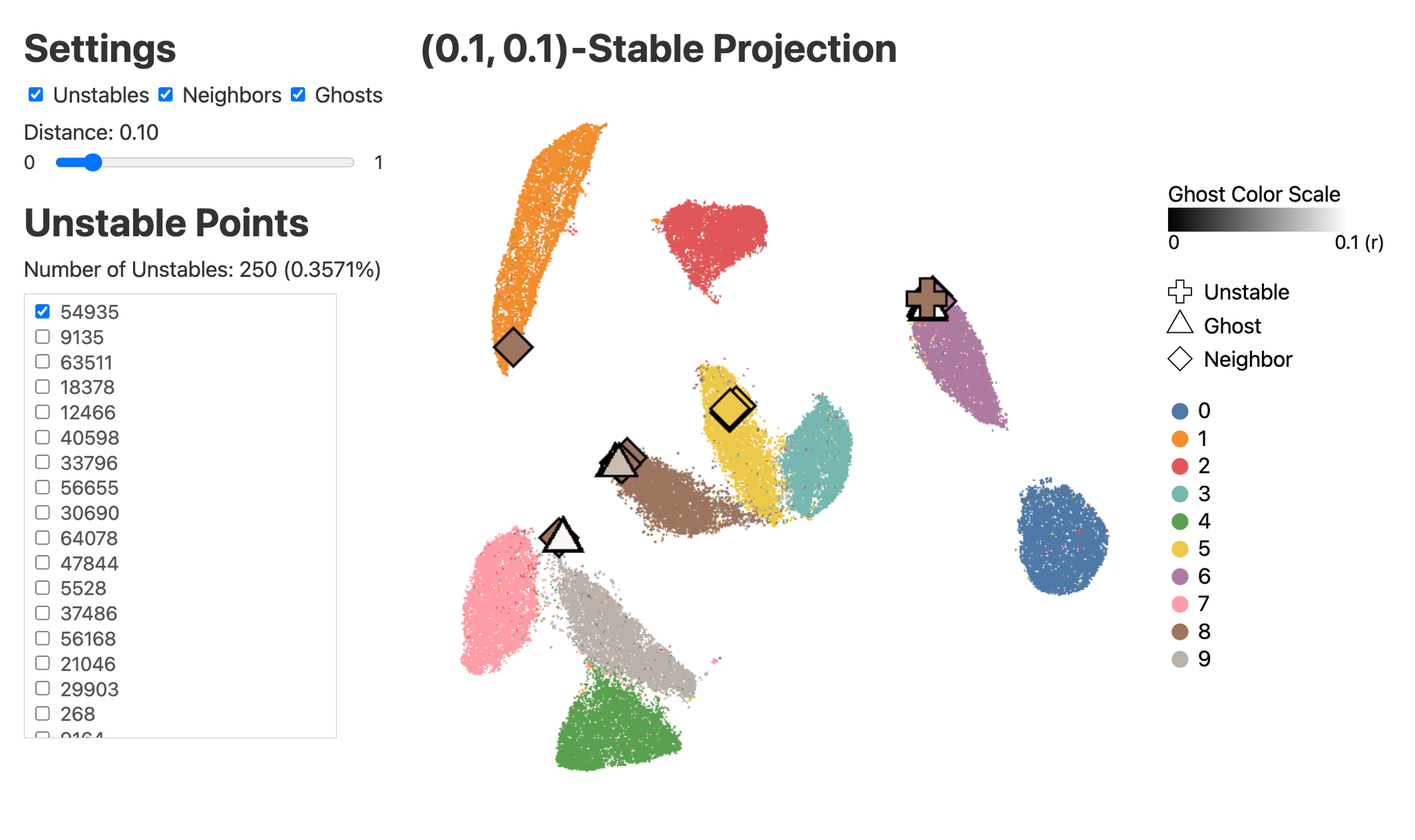}
    \caption{
    GhostExplorer, a visual interface for \gutwo, shows original and ghost projections on a zoomable scatterplot using color and symbol encodings. The control panel allows users to adjust the stability threshold $d$ interactively.
    }
    \label{fig:ghost_explorer}
\end{figure}

To facilitate the interpretation of ghost projections, we also present \ghostexplorer, an interactive visualization tool for \gutwo (\cref{fig:ghost_explorer}), implemented as a Jupyter Notebook widget and accelerated by WebGPU.
It can be invoked from a \gutwo instance, e.g., \texttt{gmap.visualize()} (Line 16 in \cref{code:ghostumap_example}).

\begin{figure}
    \centering
    \includegraphics[width=\linewidth]{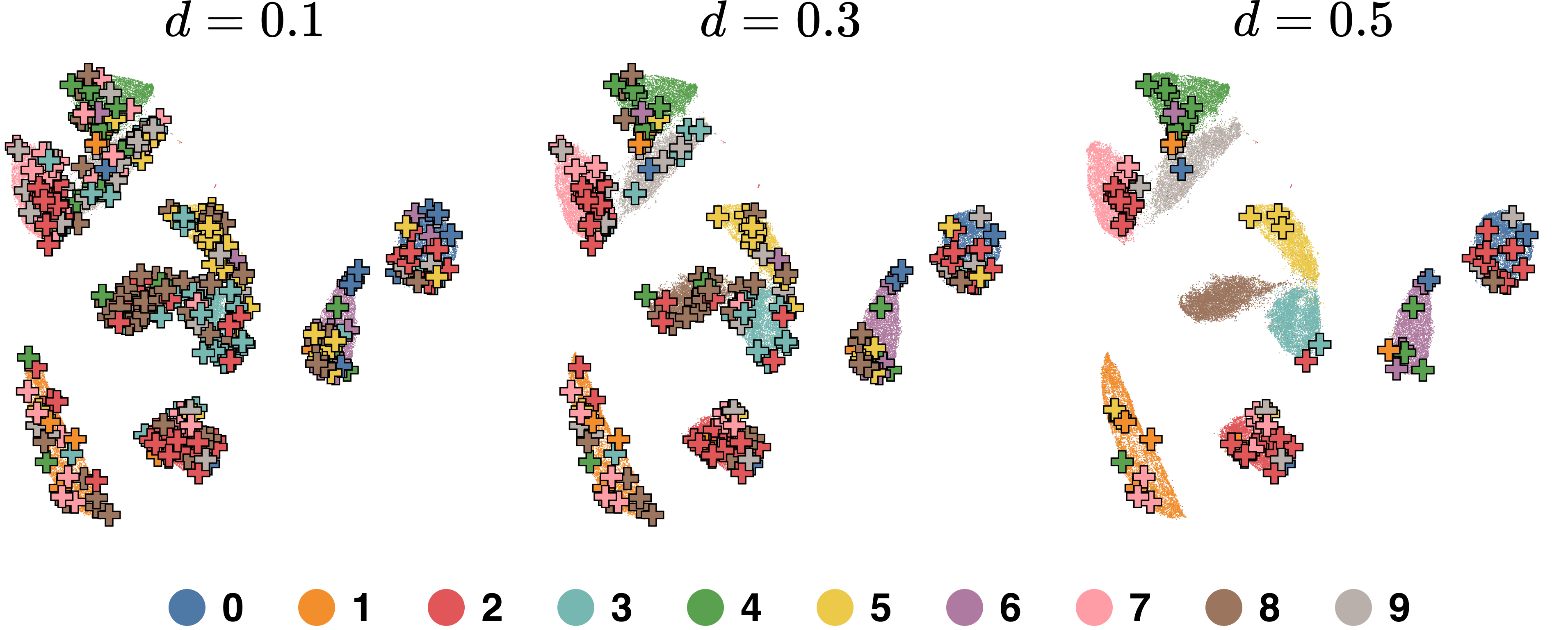}
    \caption{
    Points identified as unstable with different stability thresholds $d$. More points are identified as unstable for a small, thus strict, threshold.
    }
    \label{fig:different_ds}
\end{figure}

\ghostexplorer visualizes the projections of the original points and ghosts on a zoomable scatterplot (\cref{fig:ghost_explorer}), with optional color encoding for data labels (if available).
The user can adjust $d$ using a slider in the control panel on the left. \cref{fig:different_ds} illustrates identified unstable points for different values of $d$.
The control panel also lists unstable data points, i.e., data points whose $d_i > d$.

To inspect individual ghosts of a specific target point, the user can click on a point in the scatterplot or the checkbox of a point in the list.
This highlights the target point and its ghost projections in the scatterplot using larger symbols: a cross for the original projection, triangles for the ghost projections, and diamonds for the projections of its neighbors in the high-dimensional space (see \cref{fig:usecase}).
Ghost projections can be further color-encoded based on their distance from the original projection in the initial projection; we used lightness for this optional encoding to avoid conflict with hues often used for encoding class labels.
The user can also toggle the visibility of unstable points through the visibility checkboxes at the top of the control panel, leaving only \rd-stable points in the scatterplot.
\mw{
In the supplementary material, we present examples of analytic tasks that \ghostexplorer supports.
}
\mw{
While \ghostexplorer benefits from WebGPU acceleration, it remains subject to visual scalability issues inherent to scatterplots such as overplotting. 
Thus, an interesting extension would be to employ more scalable idioms as proposed in prior studies~\cite{mayorga2013splatterplots, jo2018declarative, lu2024visual}.
}

\begin{table*}[t]
\centering
\renewcommand{\arraystretch}{1.2}

\caption{
Benchmark results of different methods across the target datasets.
In terms of runtime, our new adaptive dropping method achieved a speed up of 2.4x on average compared to the baseline with no ghost reduction and of 1.4 compared to the halving method.
The halving and adaptive methods exhibited similar recall scores.
In contrast, the adaptive method precisely identified unstable points, showing significantly better F1 scores than the halving method.
For F1 and recall scores, we report the standard deviations in parentheses. \vspace{-2mm} 
}

\resizebox{\textwidth}{!}{
\begin{tabular}{lrrrrrrrr}
\toprule
 &  \multicolumn{4}{c}{Completion time in seconds (Speedup by ghost reduction)} & \multicolumn{2}{c}{F1} & \multicolumn{2}{c}{Recall} \\
 \cmidrule(lr){2-5} \cmidrule(lr){6-7} \cmidrule(lr){8-9}
 & & GhostUMAP2  & \gu & \gutwo & \gu & \gutwo & \gu & \gutwo \\
Dataset & UMAP  & (No reduction) & (Halving) & (Adaptive) & (Halving) & (Adaptive) & (Halving) & (Adaptive) \\
\midrule
C. elegans & 3.19 & 32.58 & 20.65 (1.58) & \textbf{14.58 (2.23)} & 0.08 \small{($\pm$0.01)} & \textbf{0.89} \small{($\pm$0.05)} & 0.87 \small{($\pm$0.04)} & \textbf{0.89} \small{($\pm$0.04)} \\

Optical Recognition & 2.90 & 27.49 & 17.75 (1.55) & \textbf{12.61 (2.18)} & 0.05 \small{($\pm$0.01)} & \textbf{0.95} \small{($\pm$0.06)} & \textbf{0.94} \small{($\pm$0.06)} & 0.91 \small{($\pm$0.10)} \\

MNIST & 10.18 & 130.15 & 78.30 (1.66) & \textbf{51.71 (2.52)} & 0.10 \small{($\pm$0.00)} & \textbf{0.94} \small{($\pm$0.01)} & 0.94 \small{($\pm$0.02)} & \textbf{0.95} \small{($\pm$0.02)} \\

Fashion-MNIST & 10.57 & 135.26 & 79.78 (1.70) & \textbf{54.29 (2.49)} & 0.04 \small{($\pm$0.00)} & \textbf{0.89} \small{($\pm$0.02)} & 0.82 \small{($\pm$0.02)} & \textbf{0.85} \small{($\pm$0.02)} \\

Kuzushiji-MNIST & 10.40 & 134.39 & 80.30 (1.67) & \textbf{54.86 (2.45)} & 0.08 \small{($\pm$0.00)} & \textbf{0.92} \small{($\pm$0.01)} & 0.91 \small{($\pm$0.02)} & \textbf{0.92} \small{($\pm$0.01)} \\

20NG & 3.73 & 37.41 & 23.14 (1.62) & \textbf{17.16 (2.18)} & 0.08 \small{($\pm$0.01)} & \textbf{0.88} \small{($\pm$0.03)} & 0.85 \small{($\pm$0.03)} & 0.85 \small{($\pm$0.03)} \\

AG News & 17.30 & 223.99 & 132.09 (1.70) & \textbf{89.26 (2.51)} & 0.09 \small{($\pm$0.01)} & \textbf{0.88} \small{($\pm$0.02)} & 0.88 \small{($\pm$0.02)} & \textbf{0.90} \small{($\pm$0.01)} \\

Amazon Polarity & 57.79 & 795.45 & 452.61 (1.76) & \textbf{325.24 (2.45)} & 0.09 \small{($\pm$0.00)} & \textbf{0.87} \small{($\pm$0.01)} & \textbf{0.84} \small{($\pm$0.01)} & 0.82 \small{($\pm$0.01)} \\

\bottomrule
\end{tabular}}
\label{tab:benchmark}
\vspace{-3mm}
\end{table*}

\begin{table}[ht]

\centering

\renewcommand{\arraystretch}{1.1}
\caption{
Summary of datasets used in the benchmark study.\vspace{-2mm}}
\scriptsize

\begin{tabular}{@{\hspace{3pt}}p{3.5cm} c @{\hspace{30pt}} c@{\hspace{3pt}}}
\toprule
Dataset & Size ($N \times$dims) & Type \\
\midrule
C. elegans~\cite{packer2019lineage} & ~~6,188$\times$50 & Table \\
Optical Recognition~\cite{asuncion2007uci} & ~~5,620$\times$64 & Image \\
MNIST~\cite{mnist} & ~~70,000$\times$784 & Image \\
Fashion-MNIST~\cite{xiao2017fashion} & ~~70,000$\times$784 & Image \\
Kuzushiji-MNIST~\cite{clanuwat2018deep} & ~~70,000$\times$784 & Image \\
20NG~\cite{20news} & ~~18,846$\times$768 & Text \\
AG News~\cite{zhang2015character} & 120,000$\times$768 & Text \\
Amazon Polarity~\cite{zhang2015character} & 400,000$\times$768 & Text \\
\bottomrule
\end{tabular}
\label{tab:datasets}
\vspace{-3mm}
\end{table}

\section{Quantitative Experiments}

In this section, we perform two quantitative experiments.
The experiments had two primary objectives: (1) to evaluate the overhead introduced by \gutwo and assess how the adaptive dropping scheme mitigates this overhead while maintaining accuracy, and (2) to analyze the impact of hyperparameters on both runtime and accuracy.

\subsection{Performance Benchmark}\label{sect:perf_bench}

The benchmark study was conducted to assess the effectiveness of the adaptive dropping scheme compared to the halving scheme~\cite{jung2024ghostumap}.

\textbf{Datasets.}
We used eight datasets, with varying sizes and data types, such as tables, text, and images (\cref{tab:datasets}), that were widely used in evaluating DR techniques~\cite{jeon2022uniform, tang2016visualizing, jung2024ghostumap}.

\textbf{Methods and Hyperparameters.}
We compared four methods: UMAP, \gutwo without any ghost reduction method (hereafter, no reduction), with the successive halving (hereafter, halving), and with the adaptive dropping (hereafter, adaptive).
For the UMAP hyperparameters for all methods, we used the default values specified in the official documentation~\cite{mcinnes2018umap}: $n\_neighbors = 15$, $min\_dist = 0.1$, and Spectral Embedding for initialization.
For \gu and \gutwo, the number of ghosts ($M$) was set to 16.
For \gu with halving, the schedule hyperparameter was set to [50, 100, 150] as proposed in the previous work~\cite{jung2024ghostumap}.
For \gutwo with the adaptive dropping, the hyperparameters were set to the default values: $lazy\_gen=0.2$, $drop\_start=0.4$, $\beta=0.2$, and $sensitivity=0.9$. 

Note that UMAP sets the number of epochs ($n\_epochs$) based on the dataset size: 200 epochs for datasets with over 10,000 data points and 500 epochs for smaller ones.
As both \lazygen and \dropstart are multiplied to $n\_epochs$, the epoch when the ghosts were generated (\lazygen) and when dropping started (\dropstart) varied depending on the dataset size.
For example, when \lazygen = 0.2, the ghosts were generated at the 100th epoch for the MNIST dataset ($500 \cdot 0.2 = 100$).

\textbf{Measures.}
We measured two metrics: the runtime (for all methods) and accuracy (for the halving and adaptive methods).
We measured the runtime to inspect the computational overhead the ghost projections imposed. 
We did not include the runtime to produce the weighted $k$NN graph as it was common for all methods.

For the halving and adaptive methods, we measured the \fone to investigate to what extent they maintain unstable points throughout the halving or adaptive dropping operations, compared to the case without these reduction operations.
To compute the \fone, we need to define the ground truth and predicted sets of unstable points.
We defined the ground truth set as the points that did not satisfy \rd-stability (i.e., unstable) when no ghost reduction was applied.

For the adaptive method, we set the predicted set of unstable points as the points that have ghsots, i.e., survived after the adaptive dropping operations, and whose $d_i > d$.
Throughout the benchmark, we used $r=0.1$ for ghost perturbation and $d=0.1$ as the threshold for unstable points.
\mw{In contrast, }the halving method does not explicitly produce a predicted set of unstable points~\cite{jung2024ghostumap}.
\mw{
Instead, it simply reduces the number of ghosts by dropping the most stable points at a given time point, leaving a set that is expected to contain all unstable points.
}
Therefore, for the predicted set, we used all points that survived after the three halving operations ($\frac{1}{8}$ of all points).
As a result, the predicted set for the halving method was much larger than the ground truth set (usually less than 1\% of points), resulting in poor F1 scores due to low precision scores. 
To mitigate such bias, we also report the recall scores of halving and adaptive methods.

\textbf{Setting.}
We used a desktop equipped with an AMD Ryzen 9 7900 CPU, with 3.7 GHz base clock, 12 cores, 24 threads, and 64GB of DDR5 RAM. GPUs were not used in the benchmark.

\textbf{Result and Discussion.}
The results of the benchmark are summarized in \cref{tab:benchmark}.
For runtime, \gutwo without ghost reduction introduced significant computational overhead, increasing the runtime by approximately 11.9 times on average compared to the conventional UMAP, which was closely aligned with the number of ghosts generated (i.e., $M=16$). 
\mw{
The halving method resulted in a runtime 7.2 times longer than the conventional UMAP on average, representing a 1.7 times speedup over \gutwo without ghost reduction.
The adaptive method further improved efficiency, with an average runtime 5.0 times longer than the conventional UMAP, achieving 2.4 and 1.4 times speedups over no ghost reduction and the halving method, respectively.
}
\del{The halving method demonstrated an average speedup of 1.7 times from \gutwo without ghost reduction.  
The adaptive method achieved significantly better speedups than the halving method: approximately 2.4 times faster than no ghost reduction and 1.4 times faster than the halving method.}

Both the halving and adaptive methods achieved high recall scores: 0.88 and 0.89 on average, respectively. 
The high recall score of the halving method is expected as it generated a large predicted set of unstable points.
Although the adaptive method generated a much smaller predicted set, it achieved similar recall scores to the halving method. 
However, these two methods showed significantly different F1 scores. 
The adaptive method achieved an average F1 score of 0.90, while the halving method scored 0.08, illustrating high scores for both precision and recall of the adaptive method.

Overall, the results suggest that the adaptive method not only provides substantial speedup (2.4 times on average) from the cases with no ghost reduction but also delivers a high F1 score, underscoring its efficacy in evaluating the \rd-stability of data points.

\subsection{Effect of Hyperparameters}\label{sect:eff_hparam}
\gutwo introduces five hyperparameters: the number of ghosts per point ($M=16$ by default), when to spawn ghosts ($lazy\_gen=0.2$ by default), when to start dropping ($drop\_start=0.4$ by default), the smoothing factor in the exponential moving average  ($\beta=0.2$ by default), and the sensitivity in dropping ($sensitivity = 0.9$ by default).
We set the default values of these hyperparameters through extensive grid searches on multiple datasets.

\begin{table}[t]
\centering
\caption{Effect of the number of ghosts ($M$)\vspace{-2mm}}
\scriptsize

    \begin{tabular}{rrrrrrrrrr}
    \toprule
     & \multicolumn{2}{c}{C. elegans} & \multicolumn{2}{c}{MNIST} & \multicolumn{2}{c}{AG News} \\
    \cmidrule(lr){2-3} \cmidrule(lr){4-5} \cmidrule(lr){6-7}
    $M$ & Time (s) & F1 & Time & F1 & Time & F1 \\
    \midrule
    0 & 3.19 & - & 10.18 & - & 17.30 & - \\
    8   & 9.28  & 0.87 & 32.04 & 0.91 & 52.42 & 0.84 \\
    \underline{16}  & 13.93 & \textbf{0.89} & 50.67 & \textbf{0.94} & 87.43 &\textbf{0.88} \\
    32   & 23.58 & 0.85 & 89.44 & 0.91 & 153.89 & \textbf{0.88} \\
    64  & 44.52 & \textbf{0.89} & 163.76 & \textbf{0.94} & 299.96 & 0.86 \\
    128   & 165.00 & 0.84 & 322.84 & \textbf{0.94} & 1185.41 & 0.87 \\
    \bottomrule    
    \end{tabular}
\label{tab:nghosts}
\vspace{-3mm}
\end{table}

\textbf{Dataset and Setting.}
We report the results on three datasets from the previous benchmark, \celegan, \MNIST, and \agnews, with various sizes and types; additional results on different datasets can be found in the supplementary material.
We used the same measures and settings as the previous benchmark.
\mw{That is, the ground truth for measuring \fone was set as the points whose $d_i>0.1$ when no ghost reduction was applied.}
We systematically evaluated the impact of each hyperparameter while keeping the other hyperparameters with default values. The specific values tested for each hyperparameter were as follows (default values are underlined):

\begin{itemize}[leftmargin=24pt]
    \item $M \in \{8, \underline{16}, 32, 64, 128\}$
    \item $lazy\_gen \in \{0.1, 0.15, \underline{0.2}, 0.25, 0.3\}$
    \item $drop\_start \in \{0.3, \underline{0.4}, 0.5, 0.6, 0.7\}$
    \item $\beta \in \{0.01, 0.05, 0.1, \underline{0.2}, 0.3, 0.4\}$    
    \item $sensitivity \in \{0.8, \underline{0.9}, 1.0\}$    
\end{itemize}

\textbf{Result and Discussion.}
\cref{tab:nghosts,tab:lazygen,tab:dropstart,tab:beta,tab:sensitivity} present the benchmark results, with the default values underlined for clarity.
For the number of ghosts $M$ (\cref{tab:nghosts}), we observed that the runtime increased with the value of $M$ almost proportionally.
This is expected since optimizing  $M$ ghosts per data point slows down each optimization iteration by approximately $(M+1)$ times. In contrast, no clear trend was observed regarding the \fone. However, using too few ghosts may limit the exploration of alternative projections using \ghostexplorer, potentially compromising their usefulness. Consequently, we propose $M=16$ as the default value, as it strikes a reasonable balance between runtime and usefulness.

For \lazygen (\cref{tab:lazygen}) and \dropstart (\cref{tab:dropstart}), we found their impacts on the runtime. 
The larger value for \lazygen defers ghost generation to later epochs, reducing the runtime.
In contrast, the larger value for \dropstart delays the adaptive dropping, increasing the runtime.
Large values for these parameters showed a subtle but negative impact on F1 scores.
Based on these observations, we recommend setting the default values of \lazygen and \dropstart to 0.2 and 0.4, respectively.

\begin{table}[t]
\centering
\caption{Effect of \lazygen\vspace{-2mm}}
\scriptsize
    \begin{tabular}{rrrrrrrrrr}
    \toprule
     & \multicolumn{2}{c}{C. elegans} & \multicolumn{2}{c}{MNIST} & \multicolumn{2}{c}{AG News} \\
    \cmidrule(lr){2-3} \cmidrule(lr){4-5} \cmidrule(lr){6-7}
    \lazygen & Time (s) & F1 & Time & F1 & Time & F1 \\
    \midrule
    0.10   & 18.05 & \textbf{0.89} & 64.93 & \textbf{0.94} & 113.17 & \textbf{0.88} \\
    0.15  & 15.88 & 0.76 & 58.35 & 0.93 & 101.79 & 0.86 \\
    \underline{0.20}   & 13.93 & \textbf{0.89} & 50.67 & \textbf{0.94} & 87.43 & \textbf{0.88} \\
    0.25  & 12.42 & 0.88 & 42.19 & 0.93 & 73.05  & 0.86 \\
    0.30   & 10.68 & 0.88 & 35.15 & 0.89 & 60.44  & 0.86 \\
    \bottomrule
    \end{tabular}
\label{tab:lazygen}
\vspace{-1mm}
\end{table}
\begin{table}[t]
\centering
\caption{Effect of \dropstart\vspace{-2mm}}
\scriptsize
    \begin{tabular}{rrrrrrrrrr}
    \toprule
     & \multicolumn{2}{c}{C. elegans} & \multicolumn{2}{c}{MNIST} & \multicolumn{2}{c}{AG News} \\
    \cmidrule(lr){2-3} \cmidrule(lr){4-5} \cmidrule(lr){6-7}
    \dropstart & Time (s) & F1 & Time & F1 & Time & F1 \\
    \midrule
    0.3   & 10.80 & 0.85 & 35.75 & 0.91 & 72.45 & 0.86 \\
    \underline{0.4}   & 13.93 & \textbf{0.89} & 50.67 & 0.94 & 87.43 & \textbf{0.88} \\
    0.5   & 17.79 & \textbf{0.89} & 65.93 & \textbf{0.95} & 112.61 & 0.87 \\
    0.6   & 20.87 & 0.84 & 79.94 & 0.92 & 142.44 & 0.86 \\
    0.7   & 24.86 & 0.83 & 96.27 & 0.89 & 163.54 & 0.79 \\
    \bottomrule
    \end{tabular}
\label{tab:dropstart}
\vspace{-1mm}
\end{table}

\begin{table}[t]
\centering
\caption{Effect of $\beta$\vspace{-2mm}}
\scriptsize
    \begin{tabular}{rrrrrrrrrr}
    \toprule
     & \multicolumn{2}{c}{C. elegans} & \multicolumn{2}{c}{MNIST} & \multicolumn{2}{c}{AG News} \\
    \cmidrule(lr){2-3} \cmidrule(lr){4-5} \cmidrule(lr){6-7}
    $\beta$ & Time (s) & F1 & Time & F1 & Time & F1 \\
    \midrule
    0.05  & 15.49 & 0.88 & 55.10 & \textbf{0.95} & 94.42  & \textbf{0.88} \\
    0.10   & 14.62 & 0.85 & 52.44 & 0.93 & 91.70  & 0.87 \\
    \underline{0.20}   & 13.93 & \textbf{0.89} & 50.67 & 0.94 & 87.43 & \textbf{0.88} \\
    0.30   & 13.90 & \textbf{0.89} & 49.17 & 0.94 & 84.40 & 0.83 \\
    0.40   & 13.36 & 0.81 & 48.67 & 0.91 & 84.15 & 0.85 \\
    \bottomrule
    \end{tabular}
\label{tab:beta}
\vspace{-1mm}
\end{table}

\begin{table}[t]
\centering
\caption{Effect of \sensitivity\vspace{-2mm}}
\scriptsize
    \begin{tabular}{rrrrrrrrrrr}
    \toprule
    & \multicolumn{2}{c}{C. elegans} & \multicolumn{2}{c}{MNIST} & \multicolumn{2}{c}{AG News} \\
    \cmidrule(lr){2-3} \cmidrule(lr){4-5} \cmidrule(lr){6-7}
     $sensitivity$ & Time (s) & F1 & Time & F1 & Time & F1 \\
    \midrule
    0.8  & 13.37 & 0.86 & 49.79 & 0.90 & 85.61  & 0.84 \\
    \underline{0.9}   & 13.93 & \textbf{0.89} & 50.67 & \textbf{0.94} & 87.43 & \textbf{0.88} \\
    1.0   & 14.49 & \textbf{0.89} & 51.80 & 0.92 & 89.90  & \textbf{0.88} \\
    \bottomrule
    \end{tabular}
\label{tab:sensitivity}
\vspace{-4mm}
\end{table}

For the smoothing factor $\beta$ (\cref{tab:beta}), despite its name, a larger value reduces the amount of smoothing~\cite{ema}, causing the moving average of distances to fluctuate more. This increased fluctuation in the average distance can lead to a higher rate of mistakenly dropped points, i.e., false positives. In fact, we observed that an excessively larger value of $\beta$ (e.g., 0.4) reduced runtime but produced a worse \fone. Based on these observations, we recommend a smoothing factor of 0.2 as it efficiently drops ghosts while limiting false positives.

Finally, for \sensitivity (\cref{tab:sensitivity}), a value of 0.8 resulted in the fastest runtime but decreased the \fone by approximately 0.04. 
This decrease of \fone is due to an excessive amount of data points that are considered outliers.
On the other hand, when \sensitivity is 1.0, we also see the decrease of \fone for the MNIST dataset.
We can expect that, for this dataset, the computation of $d_i$ was heavily influenced by outliers. 
We recommend a value of 0.9 for \sensitivity, as it maintains a high \fone by avoiding the influence from outliers in general.

\section{Usage Guidelines}
\label{sect:guideline}

This section provides usage guidelines for our framework, informed by the lessons learned from developing and experimenting with it.

\subsection{Considerations on the Choice of $r$ and $d$}

\begin{figure}[t]
    \centering
    \includegraphics[width=0.95\linewidth]{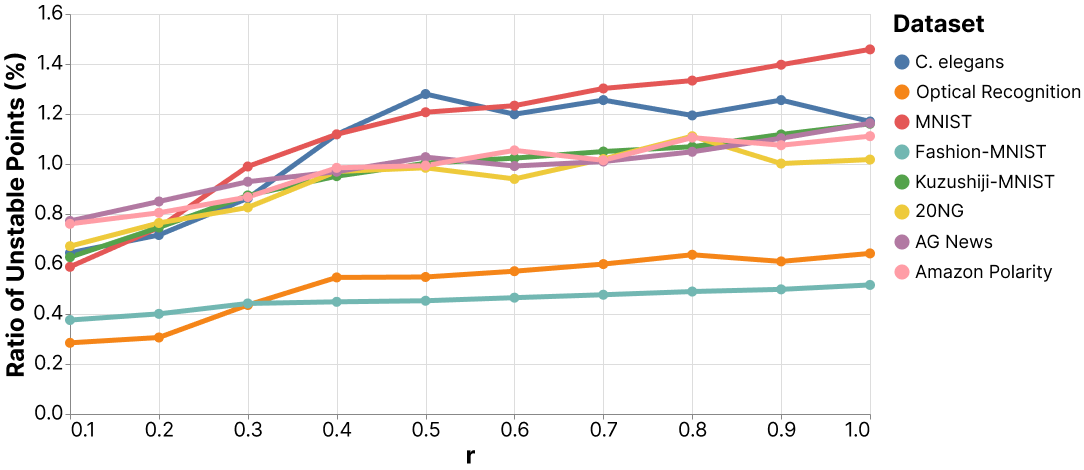}
    \vspace{-2mm}
    \caption{\edit{
    Ratio of unstable points across different values of $r$ for eight datasets in the \cref{tab:datasets}.
    As $r$ increases, the proportion of points classified as unstable (i.e., $d_i \geq 0.1$) also increases, indicating that larger perturbations tend to amplify projection variability.
    }}
    \label{fig:r_and_unstables}
    \vspace{-5mm}
\end{figure}

The hyperparameter $r$ simulates variability given in the initial projection.
As shown in~\cref{fig:r_and_unstables}, a larger $r$ results in more data points being classified as unstable when $d$ is fixed ($d=0.1$).
Another important finding regarding $r$ is that the distance between an unstable point and its ghosts in the final projection is positively correlated with $r$, but the correlation is weak---with a Pearson correlation coefficient of up to approximately 0.15.
This suggests that the hypothetical ball we assumed around the initial projection undergoes complex transformations during optimization, leading to non-uniform distortions that weaken the spatial relationship between initial and final positions.

The value of $r$ should be selected based on the intended use of \gutwo.
If the goal is to identify data points that are sensitive to stochasticity, a small value such as 0.1 (the default) is typically sufficient.
In contrast, if the focus is on isolating only the most robust points, a larger value, such as 1.0, can be used, allowing one to remove all points identified as unstable and leave the most stable points.

In contrast to $r$, the hyperparameter $d$ can be tuned without recomputation of the projection, which means that the user can interactively choose its value.
Increasing its value leads to a higher number of unstable points. 
If the goal is to remove unstable points from the projection, one can gradually increase $d$ from zero until a sufficient number of points remain for meaningful interpretation, such as identifying clusters.
Empirically, we observe that in most datasets, over 90\% of points have $d_i < 0.01$, suggesting that setting $d \approx 0.01$ retains the majority of the data.
Alternatively, if the aim is to highlight unstable points as potentially meaningful outliers, which are useful for exploring subtle cluster relationships, a higher threshold such as $d = 0.1$ may be used, which typically classifies about 1\% of points as unstable.

\subsection{Considerations on Other Hyperparameters}

For the hyperparameters of adaptive dropping, we generally recommend using the default values identified through our experiments.
If the goal is to more thoroughly explore stochastic variation, $M$, the number of ghosts (default: 16), can be increased to values, such as 32 or 64. To accelerate computation while preserving the ability to detect instability, we recommend increasing \lazygen (default: 0.2) or decreasing \dropstart (default: 0.4), rather than reducing $M$ from its default.
We found that setting either parameter to 0.3 maintains the accuracy of unstable point detection while achieving a meaningful speedup of approximately 1.3$\times$.
We generally do not recommend increasing $\beta$ or decreasing $\sensitivity$ to accelerate computation, as the resulting degradation in accuracy outweighs the speedup.

\begin{table}[t]
\centering
\caption{
Guidelines for interpreting the original and ghost projections. \vspace{-2mm}
}
\renewcommand{\arraystretch}{1.25}
\setlength{\tabcolsep}{4pt}
\small
\edit{
\begin{tabular}{@{}p{2.7cm} p{5.79cm}@{}}
\toprule
\textbf{Pattern} & \textbf{Interpretation} \\
\midrule
\textbf{P1}: Original and ghosts converge to a compact region &
\textbf{Stable.} The projection of the data point is relatively robust to stochasticity, and the location of the original can be trusted. \\
\midrule
\textbf{P2}: Original and ghosts are split into two or more distinct groups &
\textbf{Unstable.} Multiple possible projections exist; the original point represents one of the possible locations. Interpret its location with caution. \\
\midrule
\textbf{P3}: Original and ghosts are widely scattered with no clear group &
\textbf{Highly unstable.} Weak attractive forces cause the projections to be dominated by random repulsive forces. The projections are strongly subject to stochasticity. \\
\midrule
\textbf{P4}: Ghosts converge, but the original is separated from the ghost group &
\textbf{Deceptive original.} The original point may be misleading due to stochastic misplacement.
The ghost cluster may offer a more reasonable representation of the point's projection.\\
\bottomrule
\end{tabular}
}
\label{tab:interpretation}
\vspace{-4mm}
\end{table}

\begin{figure*}[t]
    \centering
    \includegraphics[width=\linewidth]{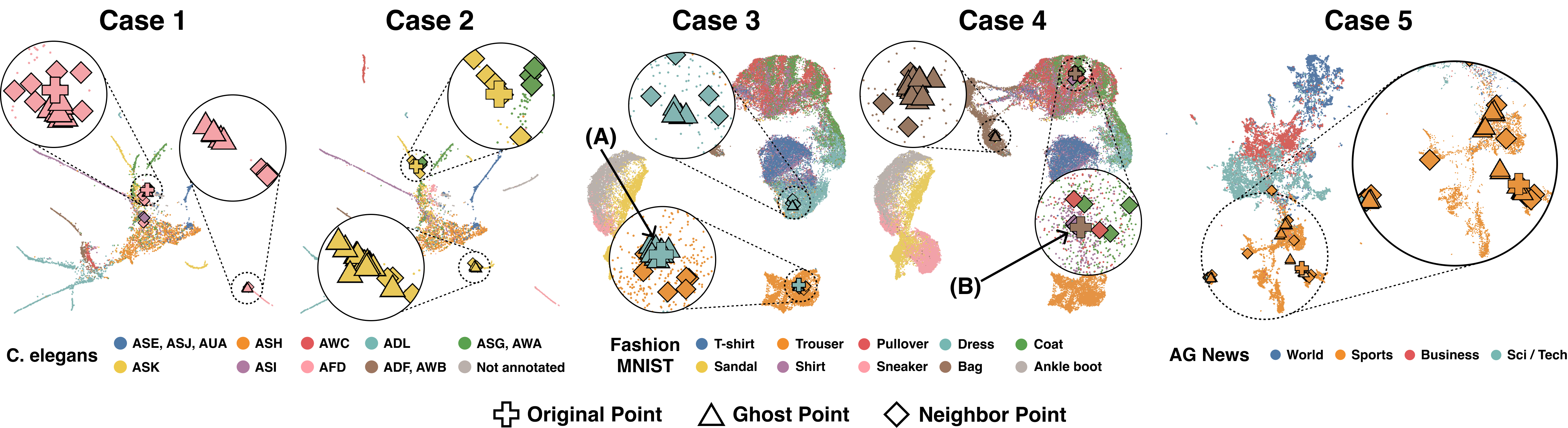}
    \vspace{-5mm}
    \caption{
    \edit{Projections of the original (cross), their ghosts (triangle), and high-dimensional neighbors (diamond) in the three datasets.
    Cases 1 and 3 show that original and ghost projections are separated into two groups.
    Cases 2 and 4 show that ghost projections are cohesive, while the original is separated from all its ghosts.
    Case 5 illustrates where both the original and its ghosts are scattered, without forming a clear group.
    }
    }
    \label{fig:usecase}
    \vspace{-3mm}
\end{figure*}

\subsection{Stability Assessment Using Ghosts}
\label{sect:ghostGuideline}

We present guidelines on assessing the stability of a data point by interpreting the distribution of its original projection and ghosts. 
We categorized common patterns we observed during experiments into four patterns, summarized in~\cref{tab:interpretation}.

\textbf{P1: Original and ghosts converge to a compact region.}
The original projection and all of its ghosts settle in a compact region, suggesting that the optimization process yields a consistent result despite stochastic perturbations and random negative sampling.
The original projection is considered \rd-stable if the original and ghost are confined in a circle of radius $d$, and they can be interpreted with high confidence.
This is the most frequent pattern we observed in practice.

\textbf{P2: Original and ghosts are split into two or more distinct groups.}
The original projection and its ghosts do not form a single cohesive cluster but instead diverge into multiple distinct regions.
While most examples exhibit a split into two primary clusters, we also observed cases where they are dispersed across three or more clusters.
This pattern suggests that the original projection is unstable, with multiple possible locations affected by random processes.
Therefore, its position should be interpreted with caution.
If a single region must be chosen to characterize the corresponding data points, it may be reasonable to select the cluster containing the majority of their ghost projections.

\textbf{P3: Original and ghosts are widely scattered with no clear group.}
The original projection and its ghosts are dispersed without forming a clear group.
Although less common, this pattern appears frequently in projections lacking clear cluster structures.
The dispersion suggests weak attractive forces from neighboring points, allowing repulsive forces from randomly sampled negatives to exert greater influence on their positions.
As a result, the original projection becomes highly sensitive to stochastic effects, and no single location can be reliably interpreted as representative.

\textbf{P4: Ghosts converge, but the original is separated from the ghost group.}
The ghosts converge tightly to a location that is spatially distant from the original projection.
Although rare, this pattern was consistently observed across multiple datasets.
In such cases, the original projection is highly unstable and potentially misleading as its location does not reflect the effect of stochasticity if ghosts were not considered.

\section{Use Cases}
\label{sect:use_cases}
\newcommand{\fmnist}{Fashion-MNIST\xspace}


In this section, we present use cases of \gutwo, where we assess the stability of UMAP projections by identifying unstable points.
We report five representative cases (\cref{fig:usecase}) from three real-world datasets drawn from diverse domains: \celegan (biology), \fmnist (images), and \agnews (text).
Additional use cases are available in the supplementary material.
For all datasets, we used the default hyperparameters and set $r$ and $d$ to 0.1.


The \celegan dataset contains transcriptome profiles of 6,188 single cells collected during \celegan embryogenesis, with each profile in 50 dimensions.
We specifically include the \celegan dataset to replicate an analysis workflow commonly used in single-cell transcriptomics.
In such analyses~\cite{packer2019lineage}, UMAP projections are commonly used for clustering cells and assigning cell type annotations to unlabeled cells using known marker genes.
However, unstable projections of data points can result in incorrect annotations of cells, thereby compromising the validity of downstream interpretations.

The \fmnist dataset consists of 70,000 grayscale $28\times28$ images of 10 classes, such as T-shirts, trousers, and sneakers.
We selected this image dataset because it allows for direct visual inspection of samples, helping to interpret the sources of projection instability.

The \agnews dataset is a collection of news articles categorized into four labels: World, Sports, Business, and Science/Technology.
We sampled 5,000 articles per category and embedded them into 768-dimensional vectors using the DistilBERT model~\cite{sanh2019distilbert}.
This dataset was chosen as a representative of text data and is notable for lacking clear separability among categories in the embedding space.


The five use cases are shown in \cref{fig:usecase}, all captured in \ghostexplorer.
We show the original projection of an unstable data point that we want to discuss as a cross, its ghosts as triangles, and its neighbors in the original high-dimensional space as diamonds.

Cases 1 and 3 in \cref{fig:usecase} show unstable points in the \celegan and \fmnist datasets, respectively, where the original and ghosts are split into two distinct regions (\textbf{P2} in \cref{tab:interpretation}).
In Case 1, six ghost points are located along the lower-right branch dominated by the ``AFD'' cell type, while the original point and the remaining ghosts are positioned near the center of the projection, where cluster boundaries are less clear.
Similarly, in Case 3, the original point is located in the ``Trouser'' cluster, while its ghosts and neighbor points split into groups aligned with both ``Trouser'' and ``Dress.''
These cases illustrate how cluster membership can be misleading for unstable points, as their projected positions may result from stochasticity rather than meaningful structure.

\begin{figure}[t]
    \centering
    \includegraphics[width=0.93\linewidth]{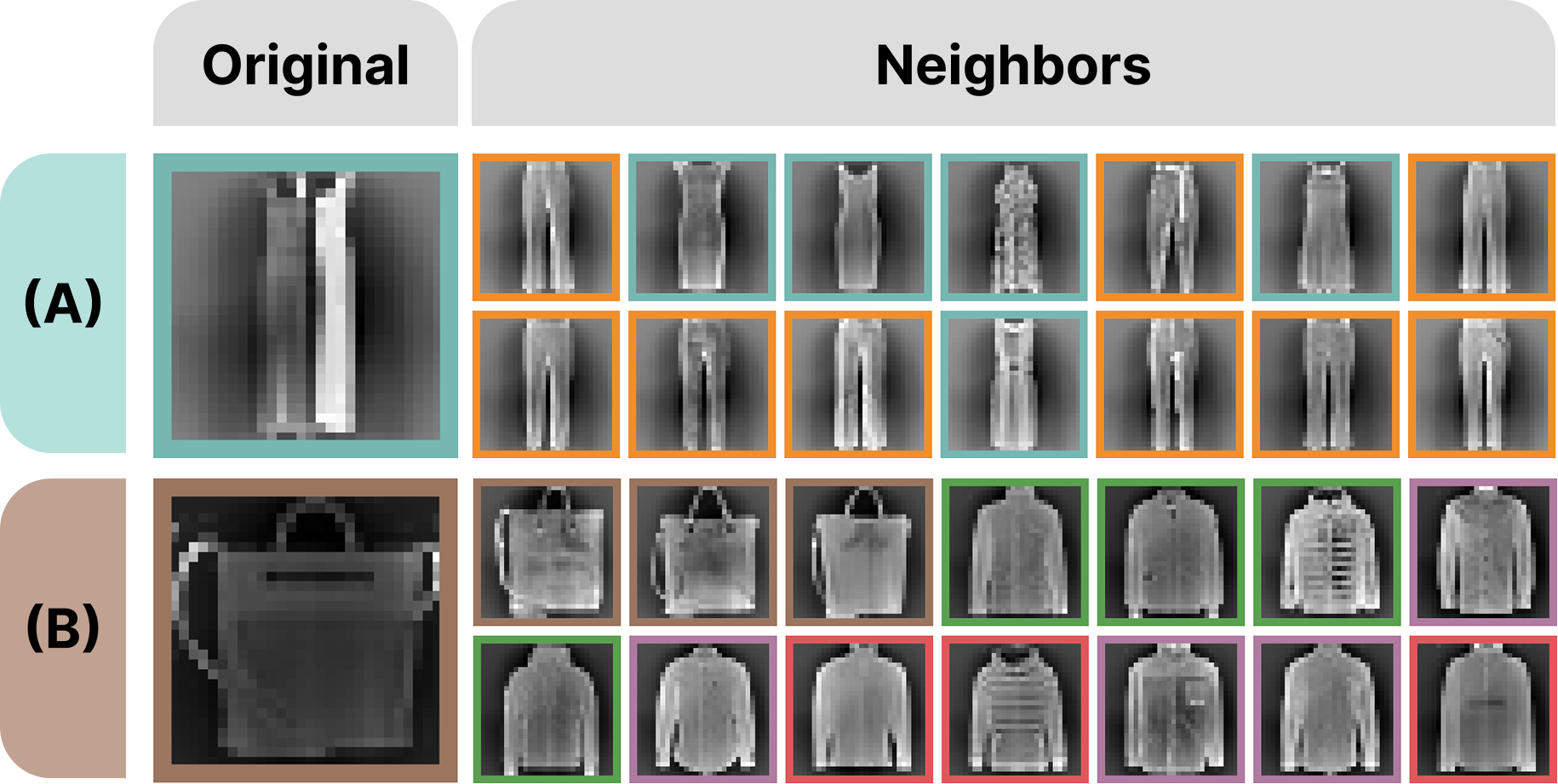}
    \caption{
    The corresponding pixel images of two data points (leftmost column) and their high-dimensional neighbors for cases (A) and (B) in \cref{fig:usecase}.
    Neighbor images are arranged from top-left to bottom-right in order of increasing distance from the original projection.}
    \label{fig:pixelImage}
    \vspace{-3mm}
\end{figure}

To gain a deeper understanding of Case 3, we inspected the pixel images of the original point and its high-dimensional neighbors (\cref{fig:pixelImage}-A).
Five of its neighbors belonged to the same ``Dress'' class as point (A), while nine belonged to ``Trouser''.
The visual similarity among these neighboring points suggests that the original image is inherently ambiguous. The ghosts are pulled toward both clusters by competing attractive forces, making them particularly susceptible to stochasticity.

Cases 2 and 4 in \cref{fig:usecase} represent two unstable data points where ghosts converge, but the original is separated from the ghost group (\textbf{P4}).
In Case 2, the original projection is near the boundary of unrelated cell types, while its ghosts consistently converge into a branch dominated by the ``ASK'' (yellow) cell type.
Similarly, in Case 4, the original projection locates the data point of ``Bag'' (brown) within the cluster dominated by the ``Pullover'' (red), ``Coat'' (green), and ``Shirt'' (purple), while its ghosts converge on the correct ``Bag'' cluster.
When examining the images of the original point and its neighbors (\cref{fig:pixelImage}-B), we found that three neighbors belonged to the ``Bag'' class, while the others came from different categories.
Both cases demonstrate that the original projection may be a stochastic outcome, leading to incorrect placement, particularly for data points having inconsistent neighbors.

Finally, Case 5 in \cref{fig:usecase} illustrates a case where the original projection and ghosts are widely scattered (\textbf{P3}).
This pattern is frequently observed in datasets where there are weak cluster structures, such as the \agnews dataset.
The widely spread high-dimensional neighbors may have contributed to this case; when neighbors are dispersed, their attractive forces become diluted, allowing repulsive forces from random negative samples to dominate.

In summary, as demonstrated in Cases 1--5, \gutwo can identify unstable projections of data points.
While we partially explained those unstable points by the dispersion or inconsistency of their high-dimensional neighbors, the reverse does not always hold.
For example, stable points not explicitly discussed (e.g., \textbf{P1}) often have neighbors scattered in the projection space but remain \rd-stable; examples of such cases can be found in the supplementary material.
This highlights that \gutwo offers insights into instability that cannot be inferred from neighbor distributions alone.
It is also important to note that class information, represented by point colors and image frames in \cref{fig:usecase} and \cref{fig:pixelImage}, respectively, is typically unavailable, as dimensionality reduction is an unsupervised method.
Even without class labels, our technique can support more reliable interpretations of projections.

\section{Discussion and Limitations}
\subsection{Comparison to GhostUMAP}

This work extends a previous study, GhostUMAP~\cite{jung2024ghostumap}, by introducing several improvements for measuring instability.
First, GhostUMAP2 generalizes GhostUMAP by allowing an arbitrary amount of perturbation to the initial projections of ghost points, in addition to the perturbation applied to the original projection by UMAP.
This additional perturbation is controlled by the parameter $r$, where GhostUMAP corresponds to the special case of $r = 0$.
By introducing $r$, GhostUMAP2 incorporates the stochasticity arising from random initializations more explicitly within the framework.

Second, while GhostUMAP measured instability using the variance among the positions of the original and ghost projections, \mbox{GhostUMAP2} uses the distance from the original projection to its farthest ghost.
This change is motivated by two reasons.
On one hand, variance is conceptually misaligned with $r$, which is defined as a distance metric; measuring instability in terms of distance offers a more interpretable and consistent framework.
On the other hand, variance does not give special consideration to the original projection.
For example, in several cases (e.g., \textbf{P4}), ghost projections form a tight cluster while the original projection lies far apart. Using variance in such cases can be misleading, yielding lower variance, as it treats all projections equally without highlighting the instability of the original point.

The third improvement concerns the acceleration of ghost computations.
In previous work, we conservatively reduced the number of points by repeatedly halving the dataset, resulting in a speedup that was limited by the number of halving steps, which was inefficient.
In GhostUMAP2, however, we were inspired by the observation that unstable points tend to have larger $d_i$ values.
This insight allowed us to adaptively reduce the number of data points based on the degree of instability observed in the middle of the computation, leading to more efficient computation.

\subsection{Limitations of \gutwo}
While our evaluation and use cases demonstrated the effectiveness of \gutwo and its adaptive dropping scheme in measuring the instability of points, several challenges warrant further discussion.
The first challenge concerns how to address unstable points.
Our work focused on identifying unstable points and understanding the reasons behind their instability by analyzing their positional variations.
However, we did not explore methods to counteract this instability.
The approach in \ghostexplorer is passive: it optionally hides unstable points.
However, more active approaches could be considered.
For example, we can intervene in the optimization process to alter the projections of unstable points.
Nevertheless, it is crucial to distinguish between instability caused by stochastic artifacts of DR techniques and the intrinsic ambiguity of the data points themselves.
Data points that are intrinsically ambiguous (e.g., \cref{fig:pixelImage}-A) may not be accurately represented as a single point in the projection, even though we fix their projections.
\mw{
Investigating the structural or algorithmic causes of instability and handling the identified unstable points accordingly remain future work.
}\vspace{-2mm}

The second challenge involves leveraging findings from \gutwo to interpret the global structures of projections.
While DR techniques are commonly used to explore the global or cluster-level structure of high-dimensional data, analyzing individual ghosts may not be well-suited for this purpose.
Addressing this limitation could involve enhanced visual support, such as enabling collective visual analysis of ghost projections.
Alternatively, computational approaches could be employed, such as incorporating the ghosts of clusters during optimization.
Extending the interpretation and application of ghosts to coarser granularities would be a promising future research direction.

The final challenge is an extension to different DR techniques. While we demonstrated the application of ghosts with UMAP as a concrete example, our \rd framework can also be applied to other DR techniques that employ force simulations, such as \mw{LargeVis~\cite{tang2016visualizing} and }$t$-SNE.
Investigating pointwise stability across different DR techniques could provide a more consistent basis for analysis and a means to evaluate the stability of techniques.

\section{Conclusion}

In this work, we present \gutwo to measure and analyze the instability of point projections, extending UMAP.
Using our \rd-stability framework, we define a point as unstable if any of its ghosts, perturbed within a radius $r$ in the initial projection, is positioned farther than $d$ in the final projection.
Ghost projections are computed through a joint optimization process, treating ghosts as faithful but alternative outcomes of the original points under stochastic conditions.
To mitigate the computational cost increased by ghost tracking, we apply an adaptive dropping scheme that achieves up to 2.4× speedup.
Through the use cases on three real-world datasets, we identify four instability patterns and demonstrate that \gutwo enables a deeper understanding of stochasticity by revealing positional alternatives and potential misplacements.
By complementing conventional UMAP analysis, \gutwo \del{provides}\mw{offers} a practical tool for identifying and interpreting projection variability, promoting a more robust \del{interpretation}\mw{analysis} of low-dimensional embeddings.

\acknowledgments{%
This work was partly supported by Institute of Information \& communications Technology Planning \& Evaluation (IITP) grant funded by the Korea government (MSIT) (RS-2019-II190421, Artificial Intelligence Graduate School Program (Sungkyunkwan University)) and the National Research Foundation of Korea (NRF) grant funded by the Korea government (MSIT) (No. RS-2023-00221186).
This work has been supported in part by the Knut and Alice Wallenberg Foundation through Grant KAW 2019.0024.
The work of Takanori Fujiwara was completed when he was with Link\"{o}ping University.
}

\bibliographystyle{abbrv-doi-hyperref}

\bibliography{reference}

\end{document}